\documentclass[review]{interact}

\usepackage{epstopdf}
\usepackage[caption=false]{subfig}

\usepackage[natbibapa,nodoi]{apacite}
\theoremstyle{plain}

\theoremstyle{definition}

\theoremstyle{remark}

\begin{document}



\title {Striving for Authentic and Sustained Technology Use In the Classroom: Lessons Learned from a Longitudinal Evaluation of a Sensor-based Science Education Platform}


\author{\name{Yvonne Chua\textsuperscript{a}, 
Sankha Cooray\textsuperscript{a, b}, 
Juan Pablo Forero Cortes\textsuperscript{a},  
Paul Denny \textsuperscript{c}
Sonia Dupuch\textsuperscript{a}, 
Dawn L Garbett \textsuperscript{d}
Alaeddin Nassani\textsuperscript{a}, 
Jiashuo Cao\textsuperscript{a}, 
Hannah Qiao\textsuperscript{a,b}, 
Andrew Reis\textsuperscript{a}, 
Deviana Reis\textsuperscript{a}, 
Philipp M. Scholl\textsuperscript{a}, 
Priyashri Kamlesh Sridhar \textsuperscript{e}
Hussel Suriyaarachchi\textsuperscript{a,b}, 
Fiona Taimana\textsuperscript{a}, 
Vanessa Tang\textsuperscript{a}, 
Chamod Weerasinghe\textsuperscript{a}, 
Elliott Wen\textsuperscript{a}, 
Michelle Wu\textsuperscript{a}, 
Qin Wu\textsuperscript{a}, 
Haimo Zhang \textsuperscript{f}
and Suranga Nanayakkara\textsuperscript{a,b}\thanks{Suranga Nanayakkara. Email: suranga@ahlab.org}}
\affil{\textsuperscript{a}Augmented Human Lab, Auckland Bioengineering Institute, The University of Auckland, New Zealand; \textsuperscript{b}Augmented Human Lab, Department of Information Systems and Analytics, National University of Singapore;
\textsuperscript{c} School of Computer Science, The University of Auckland, New Zealand;
\textsuperscript{d} Faculty of Education, The University of Auckland, Auckland, New Zealand;
\textsuperscript{e}Engineering Product Development, Singapore University of Technology and Design, Singapore;
\textsuperscript{f} OPPO Research Institute, Shanghai, China
}
}

\maketitle

\begin{abstract}
Technology integration in educational settings has led to the development of novel sensor-based tools that enable students to measure and interact with their environment. Although reports from using such tools can be positive, evaluations are often conducted under controlled conditions and short timeframes. There is a need for longitudinal data collected in realistic classroom settings. However, sustained and authentic classroom use requires technology platforms to be seen by teachers as both easy to use and of value. We describe our development of a sensor-based platform to support science teaching that followed a 14-month user-centered design process.  We share insights from this design and development approach, and report findings from a 6-month large-scale evaluation involving 35 schools and 1245 students. We share lessons learnt, including that technology integration is not an educational goal per se and that technology should be a transparent tool to enable students to achieve their learning goals. 
\end{abstract}


\section{Introduction}
Being scientifically literate is a crucial 21st century skill as it influences our ability to make decisions about our personal life and our participation in society \citep{Turiman,lederman2013}. An overarching goal of science curricula globally is to ensure that all students develop sufficient scientific knowledge and confidence in science to be informed citizens in an increasingly scientific and technological society \citep{hodson2017going}. Students who are engaged and confident in doing science in schools will develop an increased understanding of the Nature of Science (NoS). However, research shows that the NoS and scientific inquiry remain limited in scope and application in schools \citep{yacoubian2021students, andrade2020vasi,gyllenpalm2022views}.

Integrating technology into science teaching can foster higher-order thinking in students, promote investigative processes and inquiry-based learning~\citep{kiira2013oecd, davies2008integrating, TEL_duval_sharples_sutherland_2017} and develop students' conceptual understanding of science and self-efficacy in inquiry skills~\citep{wu2021integrating, crawford2005confronting, calik2013effect}. These ways of thinking underpin how science knowledge is developed. Technology can have a significant role in enabling students to work as scientists and to understand that science is a process and way of learning about the world rather than a body of knowledge to be memorised.  The challenge for teachers is to realise the potential of technology to teach about science through doing science.  This requires them to be confident and competent using technology to foster students investigative skills. 

The aim of this project was to develop a teacher-friendly technology resource which makes scientific inquiry accessible for all students in an authentic and sustainable way. 
We developed a sensor-based platform to foster students’ scientific skills -- asking questions, making observations, analysing data, and making informed decisions. We recognised that teachers needed to see the value in, and feel confident with, any technology that they expected their students to use.  Thus our design focus was to develop a platform that would support teachers' implementation of new technology in the classroom and utilise sensors that are purpose-built to align with their curriculum goals. In evaluating this work, we address the following research questions:

\begin{enumerate}
\item [RQ1] How did the design of the platform and resources support teachers' implementation of new technology in the classroom?
\item [RQ2] 
To what extent do students engage with the various features of the platform and the range of sensors available? 
\end{enumerate}
We summarise our findings and discuss the lessons we learnt across the 20 months including that technology integration in educational settings is complex and fraught; that technology integration is not an educational goal per se; and that technology should be considered a transparent tool to enable students to achieve their learning goals. We suggest some of the limitations of our study and posit future work.

\section{Background}
\vspace{-10pt}
\subsection{Technology in Science Education}
Integrating technology into science education supports inquiry-based learning in a variety of ways~\citep{TEL_duval_sharples_sutherland_2017}. Interactive technologies allow students to develop scientific reasoning, formulate hypotheses, conduct experiments, collect and analyse data, and reflect on their observations and findings, all at their own pace. Examples include web-based learning platforms with ready-to-use lessons~\citep{sciencebuddies, WISE}, plug-and-play sensors ~\citep{pocketlab, Lechelt, Evan, airbit}, custom-designed applications for data collection during field trips~\citep{doingscience}, gamified virtual environments for investigating and solving problems~\citep{rivercity} and virtual laboratories for performing experiments~\citep{virtuallab}. 

Recent developments in the accessibility of low-cost hardware and sensor-based learning technologies have allowed students to interact with real-world physical phenomena. These platforms typically utilise a combination of software and hardware to create systems that facilitate interactive and tangible learning environments~\citep{schneider2015augmenting}. Davies et al. emphasise that ``proper integration of technology can increase the likelihood that students will learn in the science classroom''~\citep{davies2008integrating} because technology positively affects student motivation, creativity, and collaboration~\citep{marshall2007tangible, horn2012tangible}.

Mobile devices such as smartphones also provide numerous opportunities for supporting sensor-based learning in science classrooms. A common approach in prior work has been to create applications capable of accessing built-in sensor data to promote science inquiry investigations~\citep{firssova2014mobile, herodotou2014design, vogel2010integrating, chu2017toward}. For example, Herodotou et al. describe their development and evaluation of an Android-based application that hosts scientific investigations by allowing students to acquire and visualise data from all available sensors on a device~\citep{herodotou2014design}.  Although students showed interest in using the tool and proposed a variety of possible investigations, the evaluation was conducted in a single session of a science and technology academy with highly motivated participants, 90\% of whom were male.  The authors were thus cautious about the generalisability of their findings, and plan future work on how to sustain interest and promote long-term engagement.  Moreover, this highlights the need for large scale evaluations involving samples where gender and socioeconomic factors are more representative of those in typical school science classrooms ~\citep{quille2017insights}.

Other work claims numerous benefits result from the use of technology for developing students' conceptual understanding of science and self-efficacy in inquiry skills~\citep{wu2021integrating, crawford2005confronting, calik2013effect}.  However, several limitations are common in existing empirical studies.  These include relatively small sample sizes~\citep{Tamar}, the use of carefully controlled conditions rather than authentic classroom environments~\citep{vogel2010integrating}, and evaluations that are conducted over short timeframes~\citep{Evan}.  This latter limitation may result in novelty effects relating to short-term interest in a new technology rather than sustained benefits in practice.  To assess the benefits of tools for developing students' scientific inquiry skills, there is a clear need for large scale evaluations conducted over realistic timeframes.
One of the contributions of the current work is a longitudinal evaluation of a sensor-based science education platform conducted over 6 months and involving over 1000 students from a range of schools.

\vspace{-5pt}
\subsection{Teachers and Technology}
Of course, the acceptance of technology by teachers is pivotal for its long-term success in the classroom~\citep{ ertmer1999addressing}. Teachers may be competent in using digital tools and devices in their everyday lives but may not exhibit similar confidence and knowledge in the use of effective technology in their science classrooms. They may rely on their personal beliefs and attitudes to guide their judgements for whether or not to incorporate technology into their science classes ~\citep{chen2008teachers, czerniak1999teachers}. Researchers report that although many factors influence the initial adoption of classroom technology, teachers' perceptions of ``usefulness'' and ``ease of use'' are critical for sustained use and successful integration of technology into the classroom ~\citep{davis1989perceived, hu2003examining}.  

In their three-year study investigating the integration of inquiry-based technology in science classrooms, Davies et al. comment on the realities of using technology in the classroom and claim that ``learning does not take place simply because technology is used''~\citep{davies2008integrating}. They found that technology should seem almost invisible in the learning process – “a transparent tool” that allows for a seamless experience when introduced to students and one that effectively enables students to achieve their learning goals.  ~\citep{davies2008integrating, davis1989perceived, hu2003examining}.  Educational technology developers need to pay close attention to teachers' considerations and focus on how sensor-based learning platforms might be implemented in the classroom ~\citep{hakverdi2012exemplary, czerniak1999teachers}.In the current work, we worked closely with teachers to design and develop a tool suitable for their needs and that they perceived as easy to use and of value. 

\section{User-Centered Design Process}
In this section we describe the user-centered design process that guided the development of our novel sensor-based science education platform to support scientific inquiry. Through conceptual discussion and idea generation with students and teachers via a series of interviews and workshops, and testing low- and mid- fidelity prototypes before large-scale deployment, we developed a plug-and-play sensor system with a web platform and complementary learning materials. 
Here we detail a three-phase design process spanning 14-months and summarise the key lessons learnt in each before reporting the findings from a 6-month deployment in real classrooms involving more than 1,000 students across 35 schools.



\subsection{Phase 1: Concept Feasibility}
\begin{figure} 
    \vspace{-15pt}
    \centering
    \includegraphics[width=\textwidth]{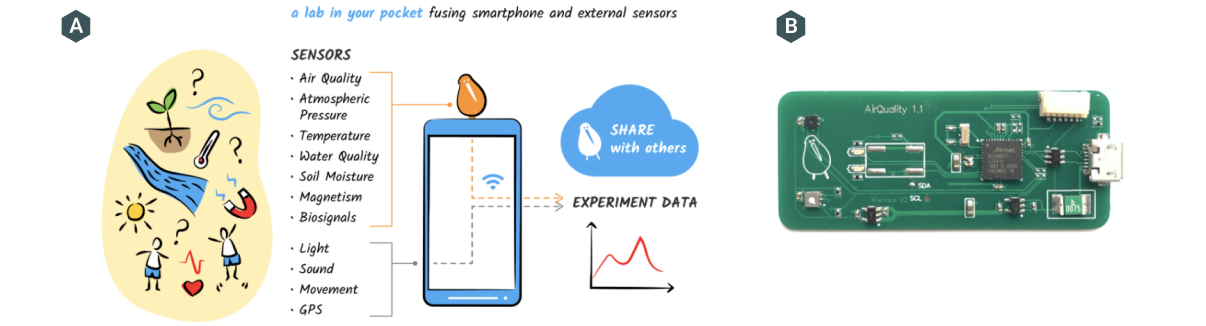}
    \vspace{-20pt}
    \caption{A) The first concept of the mobile learning platform; B) The first prototype of the plug-and-play sensor used in Phase 1.}
    \label{fig:phase1mockups}
    \vspace{-10pt}
\end{figure}

This phase involved validating our concept -- a plug-and-play sensor kit for smart phones with a complementary learning platform -- as a tool for students to conduct scientific inquiries. We created mock-ups of the concept which are illustrated in Figure~\ref{fig:phase1mockups}. 

We ran four 45-minute focus group sessions with two teachers in each group. The teachers were recruited through our personal networks from schools local to the research team. We showed them the mock-ups to gauge their receptivity to the idea and asked how they thought they might use the sensors with their students. In addition to discussing the concept, we asked the teachers how they usually ran science experiments, their observations of student engagement during the experiments, and the rules around device use in schools. Teachers were also provided a questionnaire that included a list of 10 sensors for measuring common environmental and physiological data (heart rate, UV light, brightness, sound, skin conductance, temperature, atmospheric pressure, humidity, distance to objects, water quality (pH and chlorine)). They were asked to 1) rank what they felt would be the most exciting sensors to use in class with a reason for their choice, and 2) indicate the importance of features such as ease of use, availability of guided lesson plans and alignment of those lesson plans to curricula using a 5-point Likert scale. 

\subsubsection{Findings and Learnings: }
We found that the idea of using sensors in the classroom was welcomed by the teachers. This was mainly due to their prior experience of seeing students  enjoy hands-on activities that made use of familiar contexts, and that produced an immediate response. Requirements for sensors included durability, with one teacher mentioning \textit{``replacement every 4 years is acceptable''}, and to be able to produce robust, repeatable measurements. In addition, we learnt that all of the schools had `no phone' policies but their students had access to either school-issued Chromebooks or their personal laptops during class. Therefore, we abandoned the idea of using phones and instead chose a web-based platform as it could be easily accessed via laptops. 

Teachers indicated that heart rate and UV sensors were likely to be the most exciting ones to use in class. This was due to reasons such as  students’ previous experience with the concepts, relatability to students’ life, and the relevance to the real world and the curriculum. We also learnt that generalist primary school teachers who cover all the subject areas  needed more guidance and structure to teach science. One Principal observed that such generalists \textit{``tend to lack confidence in science''}, thus indicating the need to provide complementary teaching resources alongside the sensor hardware and software platform. 

Lastly, we found that the teachers appreciated that they were engaged as part of the user-centered design process, with one stating \textit{``I am excited to be part of this''}.  They valued the fact that their expertise and opinions mattered in shaping the platform. As a consequence, all of the teachers in the initial focus groups volunteered to be involved in further work. 

\subsection{Phase 2: UV Experiment Pilot Study}
We started this phase with a prototype -- a web-based guided experiment using a UV sensor. The aim of this phase was to understand how the hardware sensor and scaffolded content was used by teachers in a classroom setting. We recruited two grade 8 classes (42 students), together with two teachers. Students participated in a 2-hour session which was led by the class teacher, with a team of our researchers present to observe and offer assistance. 

The session began with students navigating to the web-based experiment where they were asked to find the best material to protect a fictional character from UV light. 
This involved choosing the materials for investigation, formulating a hypothesis, collecting and analysing data, and discussing findings. At the end of the session, students were allowed to use the sensors to engage in free exploration of the platform.  


It was important for us to see how students interacted with our system~\citep{sanders2012convivial}, so observations were noted by the researchers present in the class. In addition, we collected student feedback on \textit{Things I liked about today's session} and \textit{Things I would change about today's session} via an online form. Both teachers were interviewed for 15-minutes for their observation of the class, thoughts on how our platform would fit into their existing curriculum, and how they might use it in the future.

\subsubsection{Findings and Learnings: }
Of the 42 students who participated, 29 described the session as \textit{``fun''}, and others used words such as \textit{``amazing''}, \textit{``awesome''} and \textit{``cool''}. We observed that some students wanted to start taking measurements immediately when the sensors were plugged in, instead of following the guided experiment procedure. Body movement and expression of the students showed that they were the most excited when measuring materials using the sensors, and appeared to be less engaged when having to discuss the results at the end. Many students wanted to investigate beyond the given experiment. Ten comments for \textit{Things I would change} were related to testing more or multiple materials. They also expressed interest in taking the sensors home. These observations and feedback affirmed that students were motivated to use the sensors, but, greater flexibility for students to engage with a variety of experiments was needed.

Both teachers were pleased with the session and thought the sensors worked well. One teacher thought the use of sensors would \textit{``allow more student driven practicals, as each would have the device to use to test their own ideas''}. The other teacher  thought that having the plug-and-play sensors without the guided content would be more useful in enabling students to perform their own experiments. This called for a reconsideration of the very structured learning material. As the teachers expressed the desire to have control over the lessons, we noted the importance of flexibility and the ease of integration with existing curricula and lesson plans.   

\subsection{Phase 3: School 
Trial}\label{sec:phase3}
\begin{figure}
\vspace{-10pt}
    \centering
    \includegraphics[width=1\textwidth]{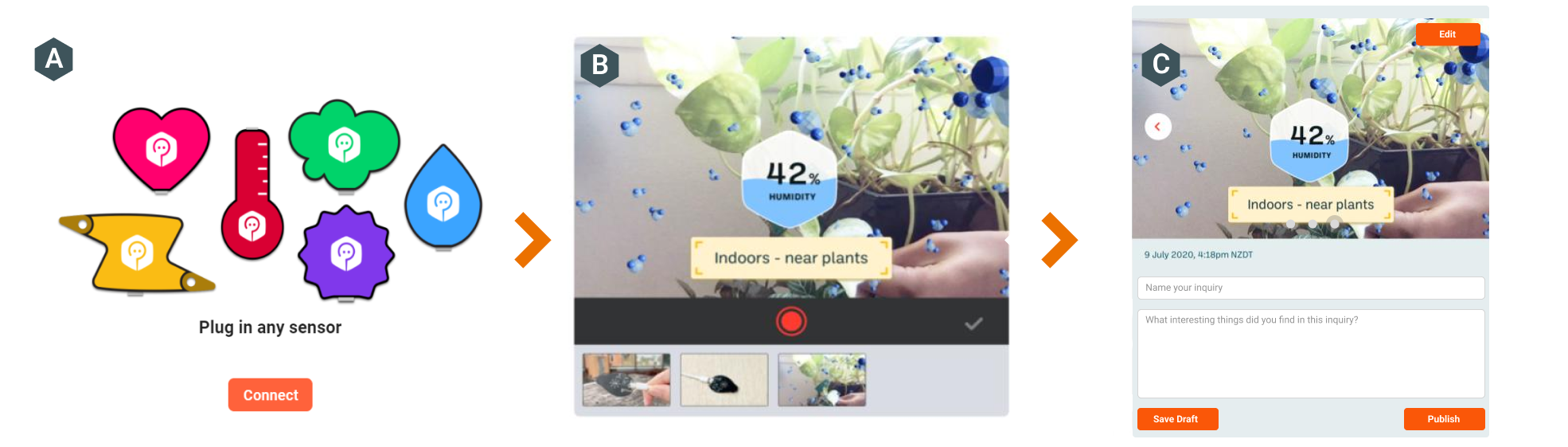}
    \vspace{-20pt}
    \caption{The procedure of publishing an inquiry. A) Plug in one of the six sensors; B) Collect up to three sets of data points; C) Name and describe the inquiry.}
    \vspace{-15pt}
    \label{fig:sensors_inquiryscreenshots}
\end{figure}

We developed a Beta version of a web platform alongside four sensors (UV, Humidity, VOC and Conductance). We used these four sensors as they were the first to be constructed. The heart rate sensor, which was rated by teachers in Phase 1 as the most exciting to use in class, was completed in time to be rolled out in the main deployment stage (see Section~\ref{implementation}). Using the platform, students were able to publish an inquiry (i.e., the investigation they were conducting using the sensors).  Figure~\ref{fig:sensors_inquiryscreenshots} shows screenshots of this process: once the sensor was connected, students could see changes in values, capture up to three data points and describe the inquiry.

We tested our system in three classrooms. A total of 79 students and 3 teachers from two schools were involved. Teachers were asked to lead the 1-hour class in which the students were able to explore the platform and use the sensors freely (see Figure~\ref{fig:classkit_classroom}b). In addition to the data collected from the server logs from the web platform, researchers were present to observe and take notes. At the end of the session students shared their feedback using post-it notes under 4 categories: 1) \textit{Three things I liked}, 2) \textit{Three things I would change or disliked}, 3) \textit{Three things I learnt}, and 4) \textit{Two things I want to try at home} (see Figure~\ref{fig:classkit_classroom}c). 

\subsubsection{Findings and Learnings: }
During the school trial, one teacher chose to give a 10-minute demonstration to their students on how to use the platform and the sensors. We recorded more inquiries published for this class that had the teacher demonstration, compared to the two classes that did not have a demonstration. We observed those students in the classes without the demonstration spent more time working out how to navigate the platform.  As a result, we noted that lesson materials should include introductory information to help teachers schedule and perform a brief demonstration.  This finding also informed how the on-boarding workshop would be run during the deployment phase.   

In terms of student feedback, 21 students liked how easy it was to use the sensors, 14 students described the session as \textit{``fun''}, and 12 students wanted more time to explore and experiment. Students liked the ability to sense or detect things that they could not see. This feedback gave us confidence that the system was engaging for students.


\subsection{Teachers' Requirements} 
\label{sec:ped_resources_research}
Teacher feedback from the Phase 1 focus groups indicated that the availability of resources was important. However the resources provided for the guided experiment in Phase 2 were, from teacher accounts, too prescriptive and specific. The amount of preparation that went into this demonstration session was unsustainable. Furthermore, it was counter to our intention for the tool to be student-driven.  With this in mind, we prepared an exemplar for each sensor and complementary learning materials with the aim to  support teachers in incorporating our system into their classrooms. 
To better understand teachers' needs, we conducted 30-minute interviews with 7 teachers, with teaching experience ranging from 2 months to 33 years. Interview questions were guided by four over-arching topics: 1) a typical science class, 2) experience in adopting new materials, 3) lesson preparation, and 4) teaching with sensors.

\subsubsection{Findings and Learnings: }
Teachers commented on the importance of being well prepared before and during teaching. They needed to understand the teaching materials and feel confident that using the technology would support their students' learning goals. If they were not comfortable using the tools, they would find it difficult to teach their students. Teachers commented \textit{``It would be helpful having someone show me how to use it first''} and \textit{``I definitely need to be confident with it first before using it with students.''} One of our priorities, therefore, was to ensure teachers were confident in using our sensors and the platform by themselves. This was addressed in the on-boarding workshops, detailed in Section~\ref{procedure}.

Most teachers favoured a guided approach to introducing the sensors. \textit{``When introducing something new, I'd introduce a very simple task''}. We provided clear instructions and an illustrative task for each sensor as a result of this feedback.

Lastly, many teachers, especially those who were inexperienced, reported that it took a lot of time to create, find, and adapt resources. \textit{``It's mostly trying to find resources for lessons, resources take up a lot of time... find the right ones to use and then typing it all up and tying it into achievement objectives.''} In addition, teachers had to ensure that the resources aligned with the national curriculum. As all the teachers relied on various web portals for their resources, providing a centralised web portal for our resources appeared be the most valuable approach for teachers.
~We describe how our implementation addresses the feedback provided by teachers in Section~\ref{sec:ped_resources}.

\begin{figure}[t]
\vspace{-20pt}
    \centering
    \includegraphics[width=\textwidth]{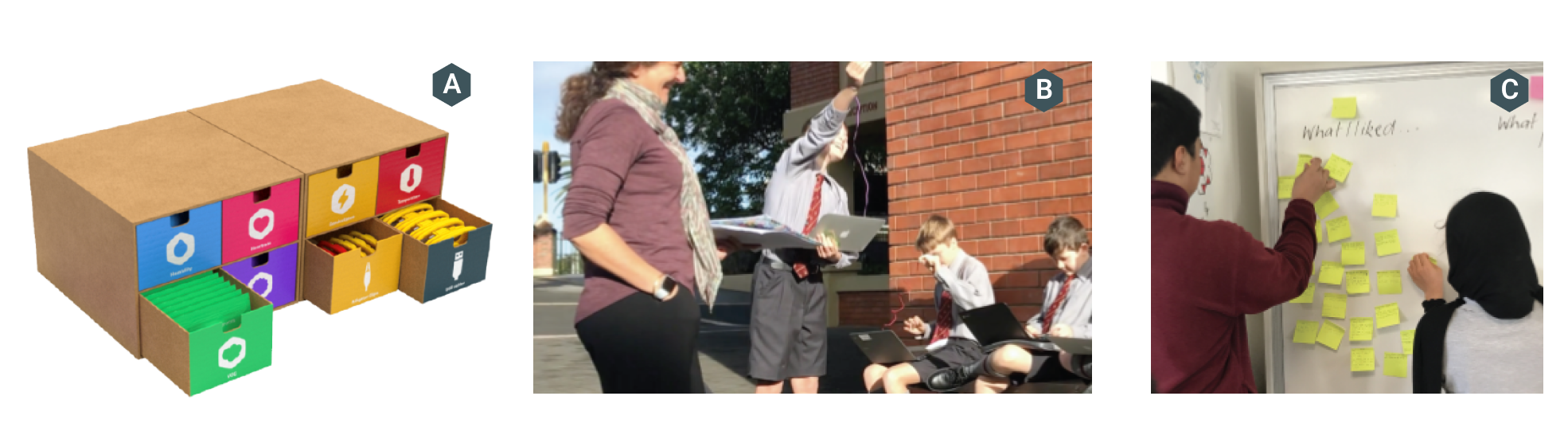}
    \vspace{-15pt}
    \caption{A) Class Kit that includes 20 each from 6 different sensors and 20 USB cables; B) Students in Phase 2 engaging with the sensors; C) Feedback session with students at the end of the session.}
        \vspace{-15pt}
    \label{fig:classkit_classroom}
\end{figure}

\section{Implementation} \label{implementation}
In this section, we describe how the final version of the learning platform  was implemented in terms of hardware sensors, web portal, and pedagogical resources. 

\subsection{Sensors}
\subsubsection{Hardware Design: }
We developed 6 different sensors (see Figure~\ref{fig:sensors_inquiryscreenshots}a) to measure the following properties: ambient temperature and humidity, ambient light and UV index, volatile organic compounds (VOC), electrical conductance, body temperature, and heart rate.
All sensors share a core design. A Microchip ATSAMD11D14  controls the sensor and handles the communication between the connected computer. A low dropout regulator provides the 3.3V supply for the microcontroller and sensor components, except for the VOC sensor which is powered at 3.1V. As we learnt from Phase 1 that the robustness and durability of the hardware was an important feature, we designed the circuit boards to also carry resettable fuses and ESD diodes to protect both the circuit board and the connected device from electrostatic events arising from mishandling. Similarly, considering the importance of creating accurate measurements, VOC, humidity, light and heart rate sensors utilise off-the-shelf digital sensors that connect to the micro-controller over an I2C bus; these components have the advantage of being calibrated by the manufacturer. The body temperature sensor is different, as its sensing component outputs an analogue voltage, thereby requiring an extra amplification and signal conditioning circuit before the analog-to-digital converter of the microcontroller. This sensor needs individual calibration due to the analog nature of the components, which was carried out prior to distribution. The electrical conductance sensor is based on a custom analogue circuit of high precision components to ensure accuracy. 

As the plug-and-play operation was an important feature, the sensors communicate and draw power from the USB port of the connected device. The micro-controller communicates using a serial communication protocol (USB-CDC(ACM)). Since this is a standard USB protocol, a majority of the modern operating systems support this communication protocol without third party driver requirements. Tested on Windows, macOS, Android, ChromeOS, and Linux based OS, the plug-and-play operation works on virtually any device with a USB port.

\subsubsection{Class Kit: }
Each class was provided with one class kit which included 20 of each sensor type and USB cables. They were packaged in easy to use cardboard drawers, and each drawer was colour-coded and clearly labelled with icons (see Figure~\ref{fig:classkit_classroom}a). Students could easily identify and choose the sensors they wanted to use during the class or teachers could access one set of sensors if they wanted students to investigate a particular phenomenon. 

\begin{figure} [t]
\vspace{-15pt}
    \centering
    \includegraphics[width=\linewidth]{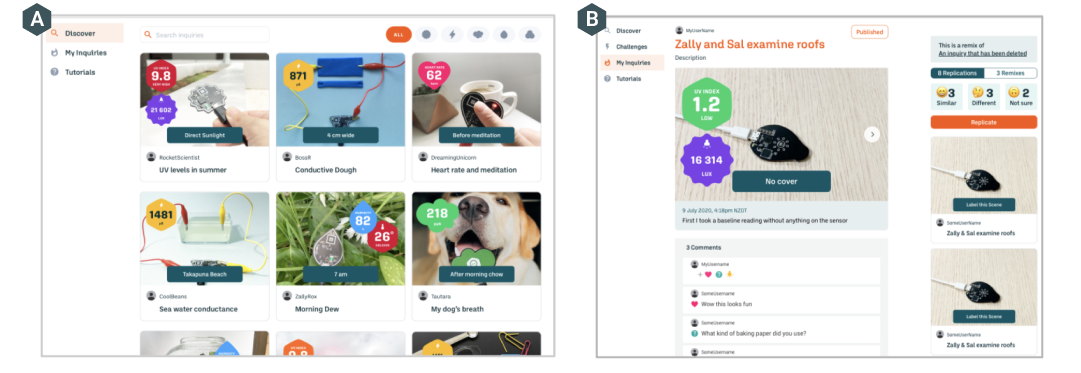}
 \vspace{-15pt}
    \caption{The Inquiry Editor. A) The Discover page shows the published inquiries, which can be filtered for each sensor; B) Sample inquiry about UV-blocking materials published by platform user.}
    \label{IE platform}
    \vspace{-15pt}
\end{figure}

\vspace{-10pt}
\subsection{Web Portal}
\subsubsection{Software: }

The web platform was built using three main components: 1) a front-end single-page app using the vue.js
~\footnote{\url{https://vuejs.org/}} 
framework, 2) a back-end using asp.net core
~\footnote{\url{https://docs.microsoft.com/en-us/aspnet/core}} web API for server-side connections and 3) a relational database to store recorded and published data using Microsoft SQL Server
~\footnote{\url{https://en.wikipedia.org/wiki/Microsoft\_SQL\_Server}}.  
The connection to sensors is established using the standard Web Serial API
\footnote{\url{https://developer.mozilla.org/en-US/docs/Web/API/Web\_Serial\_API}} 
connection via USB. 

The application is hosted on Amazon Web Services (AWS) Cloud, and the infrastructure of the deployment is described using Infrastructure as code (IaC) using Terraform\footnote{\url{https://www.terraform.io/}} which makes it easy to switch cloud providers (e.g., GCloud, AWS or Azure). Each component is built into Docker\footnote{\url{https://www.docker.com/}} containers which makes it easy to scale up and down based on usage and performance of the system using Elastic Container Services\footnote{\url{https://aws.amazon.com/ecs/}}.

\subsubsection{Workflow: }
The web portal supports any laptop running Chrome version 89 and above. The access to the web portal is orientated toward schools where teachers can create classes and invite students to join using a class code. The students can create their user account without providing any personally identifiable information (i.e., generic username and password). 

Once logged in, students can collect up to three data points, capture a photo for each set, and label and describe the inquiry in the \textit{Notes} and \textit{Description} section (See  Figure~\ref{fig:sensors_inquiryscreenshots}).  Users can choose to publish the inquiry for classmates to view or save it as a draft. Once the item is published, classmates are able to comment on the inquiry. In addition, there are two other features for users to interact with the system: \textit{replication} and \textit{remix}. \textit{Replication} is when a student produces the same inquiry as another student's, and \textit{remix} is when a student uses a published inquiry as a starting point or inspiration for their own inquiry. These options are illustrated in Figure~\ref{IE platform}.

 \begin{figure}[b]
 \vspace{-15pt}
     \centering
     \includegraphics[width=\textwidth]{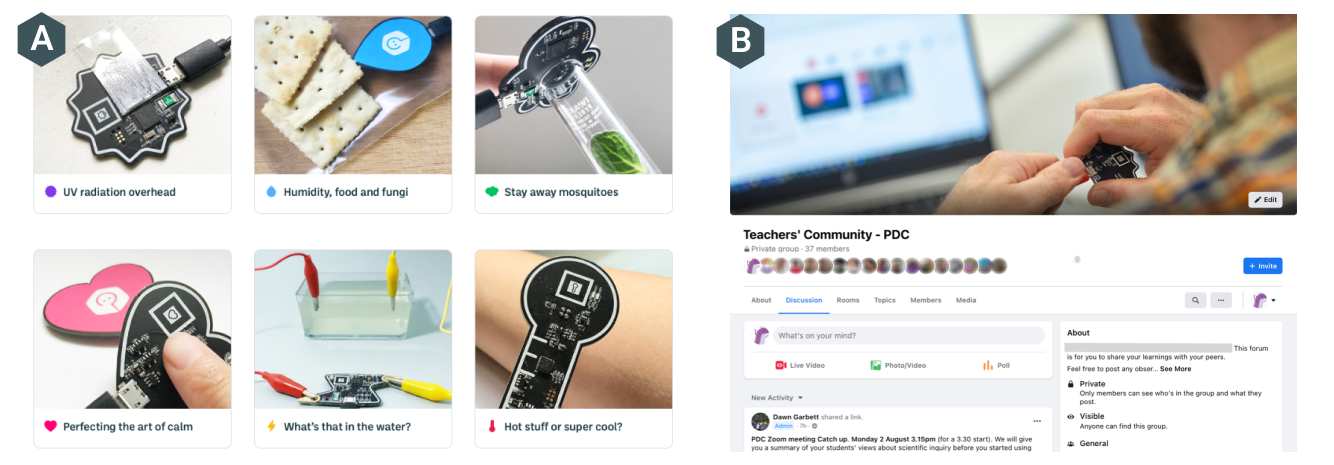}
      \vspace{-15pt}
     \caption{A) Resources and exemplar lessons aligned with the national curriculum were developed; B) A Professional Development Community was set up on Facebook for teachers to share experiences and resources with each other.}
     \label{teacher}
      \vspace{-15pt}
\end{figure}

\subsection{Teacher Resources}\label{sec:ped_resources}
We ensured that teachers felt supported and were confident to use our system, addressing concerns that had been identified in our findings  from Section~\ref{sec:ped_resources_research}, in many ways. For example, by creating a teacher resource website\footnote{Link anonymous for review} and a Facebook community group to connect teachers to one another (see Figure~\ref{teacher}). The teacher resource site was available for any educator to use, and included sensor safety instructions and video tutorials on how to use the web platform. In addition, each sensor has a complementary lesson exemplar consisting of background information, materials needed, teaching points about scientific inquiry, curriculum links, sensor information, in-class activities and extension exercises. Templates were also provided for teachers to plan their own lessons, and a Google folder was created to support sharing of resources between the educator community. In addition, our team of researchers were available at all points during the deployment for troubleshooting and added challenges and relevant material on a regular basis.

\section{Evaluation in the Real World}
To investigate the use of our system in real world classroom settings, we deployed it as part of junior science courses in 35 schools throughout \emph{\{anonymized for review\}}.
In order to address our research questions, we collected qualitative feedback from teachers via semi-structured interviews and monitored student engagement through the platform logs for 6 months.

\subsection{Participants \& Procedure}\label{procedure}
A total of 1245 students (aged between 10 and 14; 602 male, 643 female) from 35 schools participated. 66 teachers were involved, with science teaching experience ranging from less than three years to greater than 11 years. 
We first approached schools within our researchers' networks, including those schools that participated in Phase 1. 
Additional teachers were recruited via online advertisements in science teacher groups on Facebook as well as on email lists. Teachers registered their interest via a survey which asked for information related to the socio-economic status and geographic location of the school, their students' year group, number of students per class, and digital confidence. This information was used to pre-screen for the most suitable schools and teachers for deployment. The aim was to get a wide range of schools and teachers that were representative of the country's education system.

Once the teachers were selected, an in-person on-boarding workshop was conducted to teach them how to use our sensor kit and web portal. Online webinars were held for those that could not attend in person. On-boarding sessions were experiential with our team's education lead  taking the teacher's role and the participating teachers acting as students.  The aim of this, informed by the Phase 3 findings and learnings, was to allow teachers hands-on experience of our system, understanding the protocol as a student first, and then familiarising them with how they would on-board and demonstrate the system to their own students.  

After the class kits were sent and student consent forms were received, we encouraged the teachers to make use of the sensors, the web portal and pedagogical resources  as much as they could. There were no instructions given to the teachers as to which lesson material to use or how often. Our aim was to facilitate a natural integration with minimal disruption to their existing teaching plans.

\subsection{Data Collection \&  Analysis}
At the end of the 6-month deployment, semi-structured interviews were conducted to gain an understanding of the teachers' experiences and how our system was used in their classrooms. We were able to interview 13 teachers with different science specialities (e.g. biology, physics) and a wide range of teaching experience. Our interview questions had four guiding topics relating to the teacher's observations of their students' participation in class, their own experience of teaching using our system, the barriers to use or limitations of our system, and the possibility of future usage. Interview responses were transcribed and qualitatively summarised. 

We monitored interactions on the web platform and collected temporal data. Figure~\ref{fig:temporalactivity} shows the student and teacher activities over the deployment period. We received 1336 inquiry submissions from 409 students. Each inquiry published on our web platform contains \textit{title}, \textit{description} and \textit{notes}. These three parts were collectively analysed and each inquiry was given a score of Null (no science), Na\"{i}ve (minimal science), Emerging (some understanding of science) and Informed (well-informed and scientific response), adapted from Lederman's assessment of students' views on scientific inquiry~\citep{lederman2013} (refer to Figure~\ref{example_inquiries} for examples). In addition, data about which sensor was used was also logged. 

\section{Results \& Discussion}
\subsection{Teacher perception}
We now address our first research question, RQ1: How did the design of the platform and resources support teachers' implementation of new technology in the classroom?
We organised the results around the main themes that emerged from the data collected. 
\subsubsection{Teaching with sensors: } 
We noted that each teacher used the system differently in the classroom. Some teachers integrated sensors into existing teaching content. For example, the heart rate and temperature sensors were used to highlight genetic differences between individuals when teaching lessons on genetics, and the conductance sensors were used during an electricity unit. In contrast, other teachers allowed the students to play freely as they were \textit{``keen to explore the platform and see how that went for students''}. During the unguided exploration phase, one teacher identified their students' sensor preference and chose an experiment to conduct, as a class, at the end of the week; another teacher observed knowledge gaps in students' understanding in science and adapted their teaching plans accordingly to cover the unfamiliar concepts in subsequent lessons. These examples highlight the adaptability of our system.

Teachers' comments showed us that there were minimal barriers to integrating the hardware (sensors) into the classroom. Despite prior work showing that typically there is a correlation between teaching experience and technology adoption~\citep{davis1989perceived}, the inexperienced teachers in our study found our system just as easy to use as the more experienced teachers. We attribute this to the user-centered design process that engaged teachers from the beginning and allowed us to identify and preemptively address potential issues.  Our preparation of high quality resources, aligned to the curriculum and structured workshop sessions enabled confident integration of the system into the classroom.

\begin{figure}[b]
 \vspace{-15pt}
    \centering
    \includegraphics[width=\textwidth]{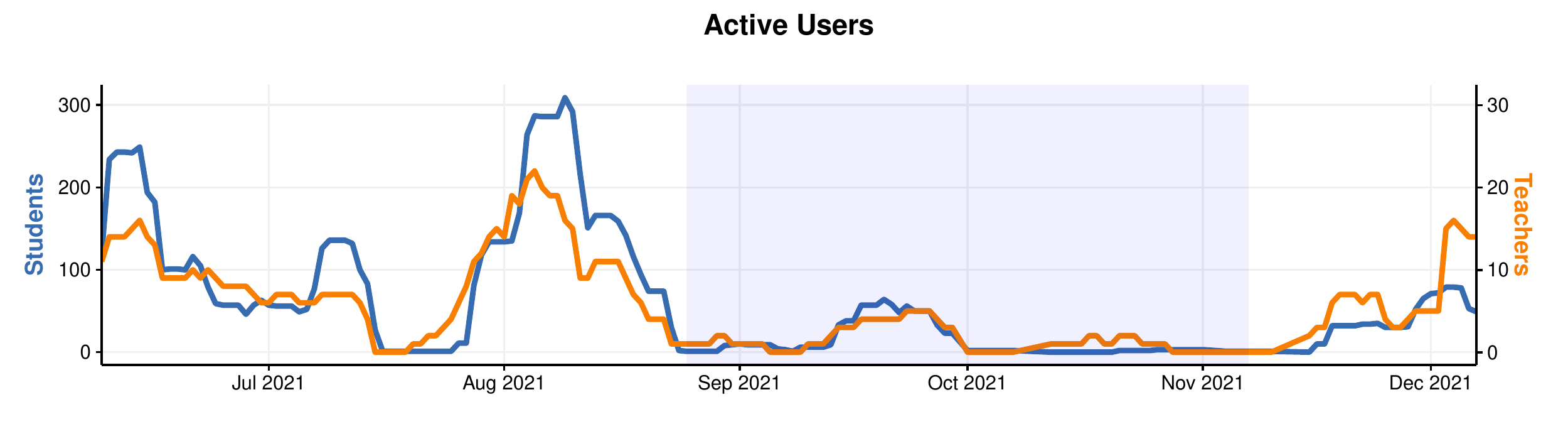}
    \vspace{-20pt}
    \caption{Usage of the platform in classrooms from June to December 2021. Drop in activity from the end of August to mid November corresponds to national lockdown due to COVID-19.}
    \label{fig:temporalactivity}
    \vspace{-10pt}
\end{figure}

\begin{figure}[t]
\vspace{-10pt}
    \centering
    \includegraphics[width=\textwidth]{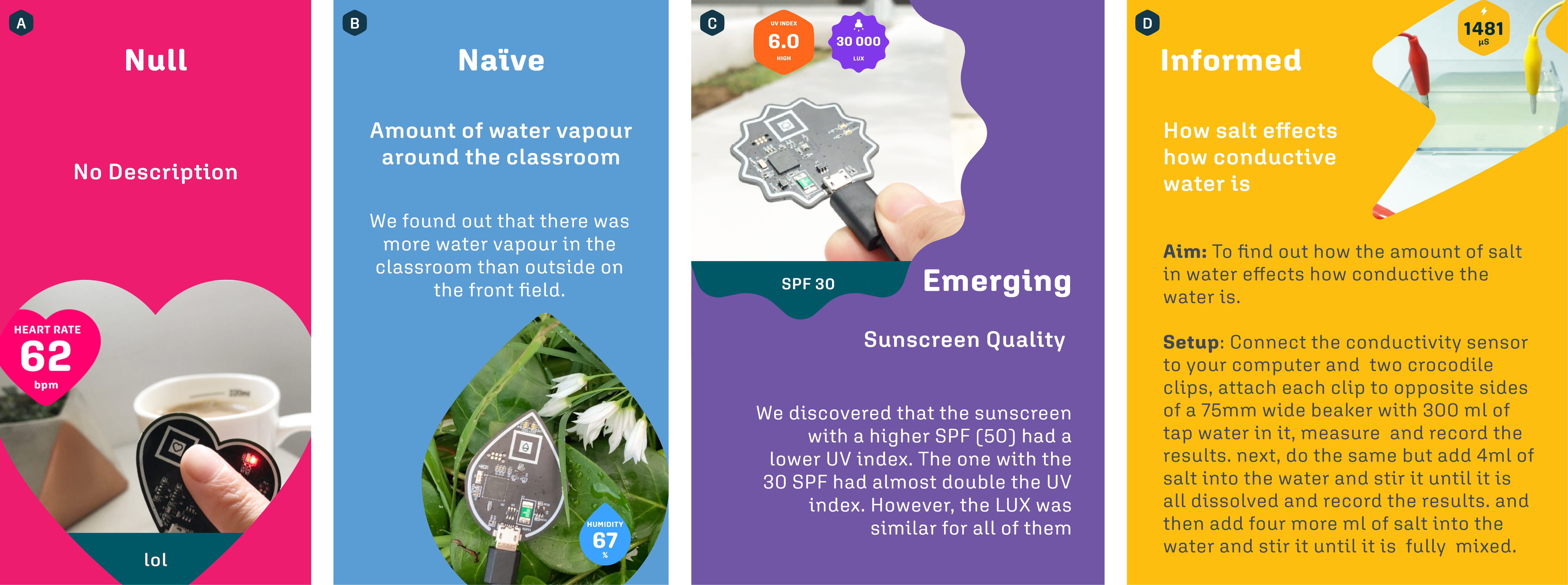}
    \caption{Representative inquiry for each score category. A) Null (no science); B) Naïve (minimal science); C) Emerging (some understanding of science); D) Informed (well-informed and scientific response).}
    \label{example_inquiries}
    \vspace{-15pt}
\end{figure}

\subsubsection{Participation and collaboration: } On observing students engagement with the sensor toolkit in class, feedback from the teachers was positive. Their observations can be broadly broken down into two categories: participation and collaboration. Most teachers reported that our sensors increased student participation in science. One teacher said ``\textit{Every single kid participated... normally there will be at least one who doesn't want to do it}''.  Similarly another teacher mentioned \textit{``some of the kids who never want to do any work, they actually did something''}. In addition, teachers liked how our system offered the potential for students to contribute in  ways that catered to their abilities. "\textit{Students who are struggling to read and write and do the math, they can make a more basic inquiry and put some basic instructions. Students who are more scientific… can have a really clear hypothesis with a title and 10 steps of instructions and all the detail that they want}".

In terms of collaboration, a few different scenarios were outlined including students working together on a single inquiry, teaching each other how to use the sensors, and giving each other instructions for inquiry replications. While one teacher noticed that some students were nervous working by themselves, the nature of our toolkit and the platform enabled every student to have a role in conducting an inquiry such as setting up the experiment, taking photos, and writing instructions. Overall, we see our system as having good potential to engage reluctant students and to create a rich learning environment for the class. 

\subsubsection{Perception of the system: } Most of the teachers interviewed highlighted the ease of use of the sensors. A few teachers emphasised the plug-and-play nature of the sensors -- ``\textit{you plug it in and get it working straight away}''. One teacher mentioned that \textit{``doing something with the sensors is much easier because you don't really have to prep. That has made it a lot more accessible.''} A frequently mentioned feature was the shape and packaging of the sensors. The teachers' observations revealed that the colourfulness of the packaging was appealing to the students, and the shape made the sensors easy to understand. Distinctive shapes and colours also allowed teachers to sort the sensors quickly for storing. 

This feedback suggests that user-friendliness was one of the principle features when thinking about the adoption of technological tools in the classroom and that our sensors fulfil teachers' requirements in this regard. However, teachers commented that the three data-point restriction the platform placed on each inquiry was too limiting (see Figure~\ref{fig:sensors_inquiryscreenshots}). As such, these teachers found their own workarounds and had their students record the sensor data on pen and paper or other software (Google sheets, docs, slides; Microsoft Excel), without publishing any inquiries or utilising other features of the web platform.

\vspace{-10pt}
\subsection{Student engagement}
Now we turn our attention to RQ2: When used over an extended period time, to what extent do students engage with the various features of the platform and the range of sensors available? 
Student engagement was primarily driven by how teachers integrated sensors into the classroom. They were given less autonomy and freedom to engage with the platform than we had hoped for as teachers, by and large, determined which sensors would be used and the context for inquiries.

\subsubsection{Analysis of Inquiries: } 
\begin{figure}[t] 
\vspace{-30pt}
    \centering
    \subfloat{%
    \includegraphics[width=0.40\linewidth]{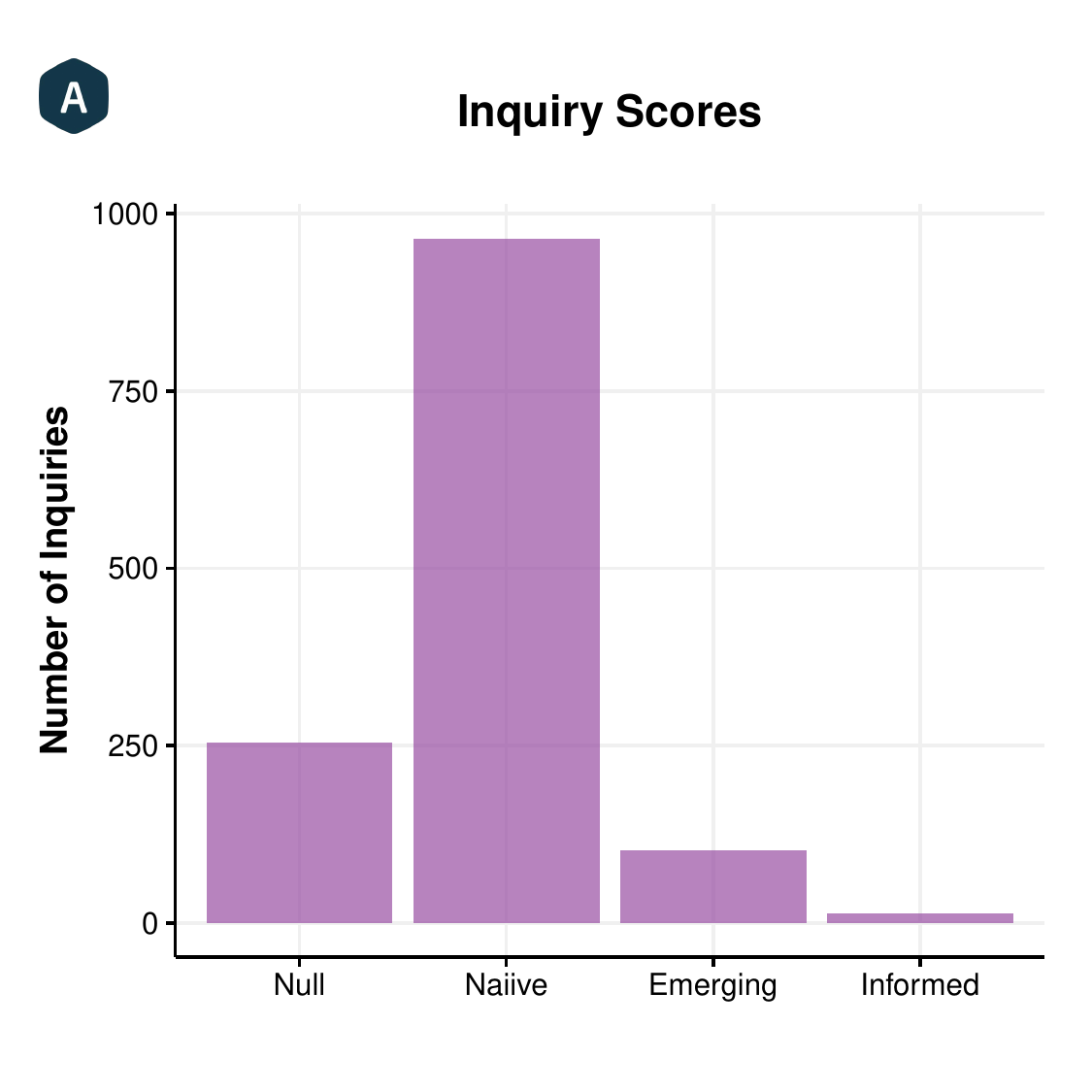}%
    }%
    \hfill%
    \subfloat{%
    \includegraphics[width=0.50\linewidth]{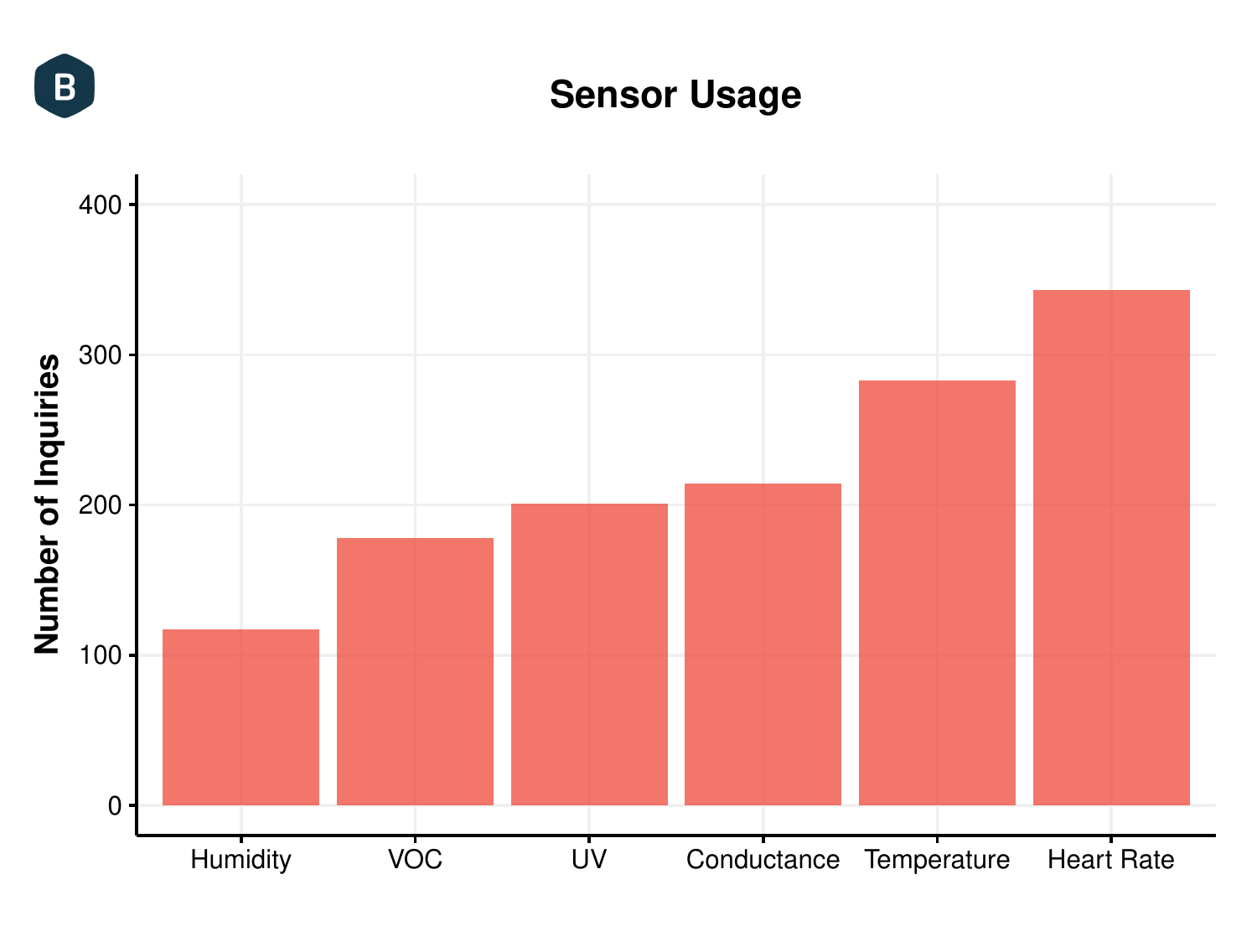}%
    }%
    \vspace{-15pt}
    \caption{A) Distribution of inquiries by scores, most inquiries scored Naïve; B) The usage of sensors.}
    \label{inquiry distribution and sensor graphs}
    \vspace{-10pt}
\end{figure}

A total of 1336 inquiries were made on the platform by 409 active users (i.e., users that made at least one inquiry). While the majority of the inquiries only contained simple descriptions of the activities, we observed a few advanced scientific inquiries which followed a scientific investigation procedure. Examples of these inquiries can be seen in Figure~\ref{example_inquiries}. The distribution of scores can be seen in Figure~\ref{inquiry distribution and sensor graphs}A. Nearly half of the 13 \textit{Informed} inquiries were from a single class. We suspect this may be due to the teacher's direction or emphasis on writing the inquiries with sufficient details for other students to replicate as discussed in `Lessons Learnt \& Limitation' (Section 7). 

\subsubsection{Feature engagement: } A total of 988 inquiries were published (where other students can see the inquiry) and 348 inquiries were saved as drafts (where only the creator can see the inquiry). One teacher thought that their students \textit{``got the feeling that they could always be doing something more''} even when they had taken measurements and added notes, they tended to not want to share work-in-progress. Despite this, many teachers saw the Publish function as a ``\textit{real plus point}''. Teachers observed that students enjoyed being able to see other students' inquiries. One teacher told us \textit{``I can hear them talking in the hallway about what they did in their class, and what other people did in their class.''}

Seven remixes and 74 replications were recorded, of which 49 were based on inquiries created by other students (60.49\%), 24 on exemplar inquiries we created (29.63\%), and eight on their own inquiry (9.88\%). Providing adequate detail for someone else to be able to replicate their inquiry had been an important aspect of the Nature of Science that we we had wanted to encourage through the platform. It seemed to us that remixing and replication were under-utilised. In fact, nearly 88\% of them were recorded from just four classes. This suggests that remixing and replication is more teacher-driven rather than student-driven and that time may be a constraint. Indeed, a few teachers responded with ``\textit{didn't have time}'' when asked about their experience with \textit{remix} and \textit{replication}. Many felt that they ``\textit{did not use the platform to its full potential}'', and they expressed interest in wanting to ``\textit{explore it [the platform] more and see what else it can do, spend more time and look at it}''.  The closure of schools across the country due to the COVID-19 pandemic during part of our evaluation exacerbated the pressure of time further.  Evidence of this disruption is visible in Figure~\ref{fig:temporalactivity}, which shows a reduction in activity over the duration of a nationwide lockdown between August and November.

\subsubsection{Sensor preferences: } 
\vspace{-5pt}
Sensor usage was analysed (see Figure~\ref{inquiry distribution and sensor graphs}B). 
In Phase 1, teachers believed that the heart rate and UV sensors would be the most exciting ones to use. 
Indeed, the heart rate sensor was the most frequently used sensor (336 inquiries). However, the second most frequently used sensor was the temperature sensor (275 inquiries). Humidity and VOC sensors were the least used sensors. One teacher mentioned that the \textit{``humidity sensor was interesting but [I'm] not sure how to use it to run a practical''}. To collect significant data with the UV sensor, students either needed a UV light source within the class (such as a UV lamp) or the ability to go out of the classroom. It appears that the sensors that engaged students the most were the ones where they could more easily record data with inquiries based in the classroom (e.g. conductance of milk with varying fat content) or which took measurements that related to their own bodies (e.g. heart rate, temperature).

\vspace{-10pt}
\section{ Lessons Learnt \& Limitations}
\vspace{-10pt}
\subsubsection*{Technology integration in educational settings is dynamic and fraught:} Developers and end-users of digital technologies do not always know, nor can they always predict, trends and applications of technologies. Moreover, due to the opaqueness of design and presentation of digital technologies, those who use digital technologies may not always understand the inner workings of the software and devices they use ~\citep{hamilton_substitution_2016}.  It is also true that developers are not often privy to the complex environmental constraints of working in a classroom with 30 students and 1 teacher as end-users. The contexts, students, pedagogical choices, as well as teachers’ beliefs and motivations all add to the complexity of navigating this space to design and implement new educational technology. This was evident when we acknowledged the teachers’ professional judgement about how to best introduce the learning platform to their classes.  The decision whether to give students a 10-minute structured introduction on how to connect the sensors and take measurements or whether to encourage students to play with the sensors and discover the capabilities was not ours to make.  Undeniably, developers need to work in concert with educators if they are to design educationally sound solutions that are fit for purpose. For any technology to be successfully adapted by a teacher, it must be designed to enhance the teacher’s self-efficacy. That is, technology should give teachers’ confidence that the technology will enhance their students’ learning. Focusing on that purpose led to our second lesson. 

\subsubsection*{Technology integration is not an educational goal per se:} This is to say that integrating technology does not necessarily lead to enhanced learning. Teachers’ decisions what, and whether, to integrate technology is a pedagogical decision reflecting the dynamic and fluid nature of teaching and learning. The important focus is on utilising technology to emphasize pedagogy and practices that support and enhance teaching and learning. Furthermore, technology can and should be used as is most appropriate to suit teaching needs. In some cases, technology may be the substitution of hard copies to online worksheets as we trialled in the carefully guided experiment in Phase 2. In Phase 3 the technology had the capacity to modify the way students did science by developing a platform which enabled students to remix and replicate one another’s science inquiries. 

\subsubsection*{Technology should be considered a ``transparent tool":} Particularly in science classrooms, technology must be easy for teachers to adapt and integrate into their regular pedagogical practices. The tools should integrate seamlessly to extend students' science capabilities -i.e. to observe, record and share data.  Rather than being a novelty and fun to play with, science tools must be seen by teachers as adding valuable learning opportunities. Science teachers look for tools that focus on assisting student learning and sustaining their engagement in challenging scientific concepts. Hamilton et al. ~\citep{hamilton_substitution_2016} affirm that the specific technological tool is not as important as how the tool is used to improve student outcomes.

\subsubsection*{Limitations in deploying technology in classrooms:} While all of the teachers were positive about using the science education platform at the end of the workshops, not all of them made use of the sensors or the platform with their classes. In fact, 9 of the 35 schools who received the sensors did not record any student sessions during the 6 month evaluation period. We reached out to them and received replies from 2 teachers who explained that a class-teacher reshuffle and student changeovers prevented them from using the sensors. These 2 schools started using the sensors after the 6-month evaluation period was over, during the next school year. However, the other 7 schools remained inactive. Of the 25 schools that started using the sensors, some used the platform solely as a sensor readout. Since teachers had complete autonomy within their classrooms to use the sensors and platform how they deemed appropriate we had no control over this.

\subsubsection*{Limitations in evaluating student learning through inquiries:} The analysis of what students had written in the title, description and notes section of their published inquiry (see Section 6.2.1) did not give a complete picture of their understanding of scientific inquiry. When pilot testing the beta version of the web platform (before the School Trial in Phase 3), we observed that a student's thought process and learning might not be articulated in their published inquiry. One student published an inquiry titled `Water molecules', with no description and 3 measurements labelled `glass water', `outside' and `breath', and no notes. That inquiry was coded as `Naive'. However, an interview with the student revealed a deeper thought process behind the inquiry: ``\textit{When we breathe out, I was expecting dry air...but when we breathe out we produce more water molecules than we would breathing in.}'' The presence of a hypothesis and attempt to validate that with observations changes the coding of the inquiry to `Emerging'.

\subsubsection*{Trade-offs in design:} In Phase 2, we observed that students wanted to start taking measurements immediately, were most excited when taking measurements, and appeared least engaged when typing in discussions and conclusions. The design of the final web platform prioritised getting students easily engaged in inquiries by having them jump straight into measurement and exploration, and giving them the option to use photos and short labels to explain what they were measuring. Typing in a detailed description and adding measurement notes were optional. This worked to get students engaged - several teachers mentioned that students were attracted by the fact that little writing was required "\textit{Oh cool, the sensors, we don't have to like write anything}". However, it came with the trade-off of not supporting or scaffolding textual documentation of their inquiry or thought processes. One teacher reported frustration at "\textit{getting kids to actually write down what they're doing, which is always a problem}".

\subsubsection*{Generalisability:} 
Despite finding evidence of positive teacher receptivity and student engagement with our system in the real world, we understand that our sensor, platform and learning material design are closely aligned  to one country and its corresponding education system. Replications in other geographic regions would be valuable. In addition, our deployment was disrupted for approximately two months by the COVID-19 pandemic. Mandated school closures followed by high rates of student absenteeism meant time teaching face to face was shortened. This in turn impacted on teachers' willingness and capacity to  integrate our technology into their lessons. 

\section{Conclusion}
In this paper we described the design, development, deployment, and evaluation of a platform to support scientific inquiry in the classroom.  The user-centered design process spanned 14-months and allowed us to create a technological tool that was accepted by all teachers, regardless of their prior experience. The large-scale real-world deployment over a period of 6-months provided novel insights that would not have been seen over a short time frame or in tightly controlled classroom settings. 
Compared to typical science classes, students expressed excitement and a greater desire to participate in the learning process, including those  students who did not normally engage well. Teachers were instrumental in how and when the technology was introduced and used. We continue to support them in their integration of our tool and develop and modify it to better meet their requirements. Our future work aims to address some of the issues still unresolved. We have developed one platform which can take continuous measurements to overcome the limitation of 3 data points. Next we will develop a mobile application so that sensors can be taken into the "field" and data stored in the cloud. We are striving to provide a technological tool kit which will change the way teachers and students engage with learning about science through doing science that is relevant and authentic. 


\section{Acknowledgements}
We would like to acknowledge the all schools including teachers and students who participated in this project. Finally, we acknowledge that all authors made critical contribution at various junctures of this long complex project. Therefore, author names were simply represented in alphabetical order.

\section{Funding}
This work was supported by the Curious Minds grant by MBIE, New Zealand as well as Assistive Augmentation research grant under the Entrepreneurial Universities (EU) initiative of New Zealand. 

\section{Conflicts of Interest}
We have no conflicts of interest to declare.


\bibliographystyle{apacite}
\bibliography{proceedings}

\end{document}